\documentclass[floatfix,aps,prl,showpacs,amsmath,nofootinbib,
preprintnumbers,twocolumn]{revtex4}

\usepackage{colordvi}
\usepackage{graphicx}

\usepackage{dcolumn}
\usepackage{bm}

\begin{document}

\preprint{}

\title{Late time failure of Friedmann equation}
\author{Alessio Notari}\email{notari@hep.physics.mcgill.ca}
\affiliation{Physics Department, McGill University, 3600 University Road,
             Montr\'eal, QC, H3A 2T8, Canada}

\date{\today}

\begin{abstract} 
It is widely believed that the assumption of homogeneity is a good zero{\it th} order approximation for the expansion of our Universe. We analyze the correction due to subhorizon inhomogeneous gravitational fields.
While at early times this contribution (which may act as a negative pressure component) is perturbatively subdominant,
we show that the perturbative series is likely to diverge at redshift of order 1, due to the growth of perturbations. 
So, the homogeneous Friedmann equation can not be trusted at late times.
We suggest that the puzzling observations 
of a present acceleration of the Universe, may just be due to the unjustified use of the Friedmann equation and not to the presence of a Dark Energy component.
This would completely solve the coincidence problem.

\end{abstract}

\pacs{98.80.cq}

\maketitle

%%%%%%%%%%%%%%%%%%%%%%%%%%%%%%%%%%%%%%%%
%%%%%%%%%%%%%%%%%%%%%%%%%%%%%%%%%%%%%%%%
\section{Introduction}

Since the birth of modern cosmology the theoretical framework used to describe the evolution of the observed Universe has been
based on the Einstein equations applied to a perfectly homogeneous and isotropic model filled with various types of
fluid components.
It is universally believed, since the Universe on large scales looks almost homogeneous, that the assumption of homogeneity is a good zeroth order approximation 
for describing our Universe, and on top of this one usually adds linear fluctuations (with initial amplitude of order $10^{-5}$, as measured by CMB experiments) in order to describe the
growth of density perturbations and the formation of structures. Some steps beyond this paradigm have been made 
for describing
late time behaviour of small scale density perturbations (which at late times become non-linear) and more recently 
in the second order relativistic perturbation theory to catch new effects (as non-gaussianities in the spectrum of primordial perturbations).

However, few efforts have been made to understand how reliable is the assumption of homogeneity. 
Of course there are corrections to this description due to perturbations: the nonlinearity of Einstein equations implies that the presence of perturbations modifies the background. The usual way of thinking implies that these corrections should be small: since they are nonlinear, the effect should appear as a second order effect, so they should be of order $10^{-10}$.

However this reasoning is too naive, because it does not take into account the fact that perturbations grow during the matter era.

Also, in recent years the exploration of the universe at redshifts of order unity has
provided information about the time evolution of the expansion rate of the
universe.   Observations indicate a very unexpected fact: the universe is presently undergoing a
phase of accelerated expansion \cite{acceleratedreview}. The accelerated
expansion is usually interpreted as evidence for a negative pressure 
component to the mass-energy density of this homogeneous model universe. The nature of this ``Dark Energy'' fluid is a huge mystery.

The goal of this paper is to show that this interpretation can not be trusted, since the approximation of perfect homogeneous background becomes inadequate at late times due to the growth of perturbations and it fails at sufficiently late times.
In other words the average description of our Universe using Friedmann equation at small redshift is likely to be incorrect, due to the effect of the inhomogeneities. 

It is striking the fact that the Friedmann equation fails around the same epoch at which the acceleration is observed. So a Universe filled only with matter, may not necessarily be in contrast with a late time acceleration: it might just be due to the unjustified use of the Friedmann equation.
This would completely answer to the coincidence problem: at some time around structure formation the fluctuations become big enough to influence the background, even irrespectively of the initial amplitude of the fluctuations \footnote{We thank N.~Shuhmaher for pointing out this fact.}.

The fact that we are pointing out here is entirely due to subhorizon perturbations, and so it is completely different from what has been suggested in~\cite{KMNRp}. In that work it has been suggested that a large variance due to superhorizon modes can have an important effect on the observed expansion: if we live in a large void we might experience an expansion very different from the usual FRW.

We analyze here instead the mean effect of inhomogeities on the expansion rate at perturbative level (using the results of~\cite{KMNR}) and estimate the mean correction which is unavoidably present in any region of the Universe. This may be interpreted as the contribution of inhomogeneous gravitational fields to the energy content of the Universe. It is in fact a pure nonlinear effect, due to the fact that gravitational fields gravitate.\footnote{This effects are often called ``backreaction'' in previous literature.}. We will analyze the contribution of the subhorizon modes, ignoring the influence of the unknown superhorizon modes. We stress so that there are no free parameters here and in principle everything is calculable and predictable.

To be able to catch this nonlinear effect, the first non-trivial order is second order. At this level, we found
 that the mean contribution of the second order is ${\cal O}(10^{-5})$\footnote{The same result was found in \cite{lam}, as an order of magnitude estimate.}. It can also be shown that this contribution may act as an effective negative pressure component.
This gives rise to a first crucial observation: the effect is five orders of magnitude bigger than what is naively expected, and this is due to the growth of perturbations after matter-radiation equality.

We point out here that although this contribution may still appear small, one has to consider the full perturbative series. By a simple order of magnitude estimate we show that in most of the history of the Universe, the magnitude of higher orders is decreasing, and so the perturbative approach is likely to give a good description.
However, the main result of this paper is that the expansion parameter becomes of order $1$ at redshifts of order $z\lesssim 1$, and not $10^{-5}$ as one might naively think.
This tells us that all the terms in the perturbative series are going to be of the same order at late times. There is some uncertainty on what is exactly the redshift at which the expansion parameter becomes $1$, but it is certainly close to present time.
As we said, at the very same redshifts the Universe is observed to accelerate. We speculate that the acceleration might just be the result of the fact that the series diverges, and that the present $\Lambda CDM$ concordance model might just be regarded as an effective theory in which all the terms of the series have been resummed, giving rise to an effective negative pressure component.

\section{The mean effect of inhomogeneities}

We base our analysis on the results of~\cite{KMNR}, and we recall the notations used there. 
We consider a Universe filled only by irrotational dust and choose the
coordinates of an observer at rest with the dust ({\it i.e.,} comoving
coordinates), and with the same time coordinate for every point of the
spatial hypersurface ({\it i.e.,} synchronous coordinates). This
system of coordinates can be chosen if the Universe is filled by a
single pressureless component.

Our approach is to start with a flat matter dominated Universe (so with an initial matter energy density fraction $\Omega_M\equiv \frac{\rho_M}{\rho_C}\equiv\frac{8 \pi G \rho_M}{3 H^2}=1$, where $\rho_M$ is the matter energy density, $\rho_C$ is the critical energy density, $G$ is the Newton constant and $H$ is the Hubble parameter), and compute perturbatively the correction. However we are speculating that the correction itself drives the critical density $\rho_C$ away from $\rho_M$ due to the fact that $H$ does not evolve as in a matter dominated homogeneous universe, at late times. So, when using the expression of the matter power spectrum (shape factor and normalization) for estimating the effect, we will not use as a present value $\Omega_M^{(0)}=1$\footnote{With the superscript $^{(0)}$ we always mean present time.}, but a smaller value, as measured by observations on the turnover point of the matter power spectrum.
On the other hand, when solving Einstein equations and getting the time evolution of the growth function for density perturbations we assume, in a perturbatively consistent way, matter domination, and this will be valid until the breakdown of the approximation (that is $z$ of order 1).

We will call $\tau$ the (conformal) time in this gauge, and $x^i$ the
spatial coordinates, so that the metric has the form
\begin{equation}
ds^2=a^2(\tau) \left[ -d \tau^2 + \gamma_{i j}(\tau,x^i) dx^i dx^j  \right]
\label{metric}
\end{equation}
A prime will denote derivatives with respect to conformal time. We also introduce a proper time by $dt=a d\tau$.
The metric will be expanded perturbatively (\cite{MMB}) expressing all the metric variables as a function of a single initial condition , given by the function $\varphi(x)$, which is constant in time and related to $\delta^{(1)}$, the 
first-order density perturbation, by the Poisson equation,
\begin{equation}
\nabla^2\varphi = \frac{\kappa^2}{2} a^2\rho_M^{(0)}\delta^{(1)} .
\label{nabla2varphi}
\end{equation}
We assume initial conditions as given by inflation, neglecting initial vector and tensor modes (\cite{bartolo}).

Then we will be interested in the local expansion rate $\theta\equiv D_{\mu} u^{\mu}$ as seen by a comoving observer. We are going to average this quantity over a finite region of size $R$ at fixed time $\tau$ with the following prescription:
\begin{equation}
\langle {\cal O}\rangle(\tau)
\equiv \frac{\int d^3\!x \, \sqrt{\gamma(\tau,x^i)} \,
{\cal O}(\tau,x^i)}{\int d^3\!x \, \sqrt{\gamma(\tau,x^i)}} ,
\label{average}
\end{equation}
After having computed $\langle \theta \rangle$, we compare this with the naive expectation ($3\sqrt{\kappa^2 \frac{\langle \rho \rangle}{3}}\equiv 3H\equiv\langle \theta \rangle-\langle \delta\theta \rangle $) that an observer would have measuring the average energy density $\langle \rho \rangle$ and applying the homogeneous Friedmann equations. So what we want to estimate is the quantity:
$$
\frac{\langle\delta\theta\rangle}{3H} \equiv 
\frac{\langle\theta\rangle-3\sqrt{\frac{\kappa^2 \langle \rho \rangle}{3}}}{3H}
$$

Also, we stress that here we are interested just in the mean corrections, which means that these are effects that are present in any realization of the region of radius $R$. So this means that the extension of the region $R$ is irrelevant for this discussion. As we found in~\cite{KMNR} the mean correction is due to modes which are subhorizon, and in fact the main contribution comes from wavelenghts of ${\cal O}(20-50)$ Mpc, and it is insensitive to the infrared and to the ultraviolet. These variances around this mean values are also interesting, as they seem to depend on superhorizon modes (see \cite{KMNR},\cite{KMNRp}), but we do not discuss them here and simply assume they are negligible (this is certainly true for example if fluctuations with wavelenght bigger than the present Hubble radius are negligible).

 The result that we found consistently at second order in~\cite{KMNR} was:
\begin{eqnarray}
\hspace{-2mm} 
\frac{\langle\delta\theta\rangle}{3H} & = & \frac{1}{a^2H^2} \left[
- \frac{5}{9} \left\langle\nabla^2\varphi\right\rangle_1
+ \frac{5}{3} \left( \left\langle\varphi\nabla^2\varphi\right\rangle
+ \frac{13}{18}\left\langle\varphi_{,i}\varphi^{,i}\right\rangle \right)
+\right. \nonumber\\
&& \left.+ \frac{\tau^2}{27}\left(\left\langle\left(\nabla^2\varphi\right)^2
\right\rangle -\left\langle\varphi^{,ij}\varphi_{,ij}\right\rangle\right)
\right . \nonumber \\
& & \left. - \frac{25}{9}\left\langle\varphi\right\rangle_1
\left\langle\nabla^2\varphi\right\rangle_1
- \frac{23\tau^2}{216}\left\langle\nabla^2\varphi\right\rangle_1
\left\langle\nabla^2\varphi\right\rangle_1 \right]  \, ,   
\label{risultatoWithBt}
\end{eqnarray}
where $\langle ... \rangle_1\equiv \int (...) d^3x / (\int d^3x)$.
Let us comment on the various terms that are present here.
First, note the absence of terms that go as $\varphi^2$, without gradients.
\footnote{This was found first in~\cite{gesh}, where the expansion (without gradients) was computed in the longitudinal gauge, going then back to the comoving coordinates.}

Then, as long as we are interested only in the statistical mean, many terms are irrelevant.
Indeed the first order term has obviously zero mean, and the terms in the last line are small if we perform the spatial averages on sufficiently large scales (for example at horizon scales, they are completely irrelevant).

So, only two types of terms are potentially important: the terms with two spatial derivatives  $\left\langle\varphi\nabla^2\varphi\right\rangle
+ \frac{13}{18}\left\langle\varphi_{,i}\varphi^{,i}\right\rangle$, and the terms with four spatial derivatives.

As already noted in~\cite{Rasanen}, a term with four derivatives would give a correction of order 1 to the expansion at present times, already at second order. In fact, as we are going to show again, the presence of gradients boosts the corrections: the more gradients we have, the bigger is the correction. However, an important property to keep in mind is that the mean of a total spatial gradient is zero (this can be easily verified going to Fourier space (following~\cite{KMNR}) )
In other words, as long as we are interested only in the mean, we are allowed to derive by parts our expression.

So, doing that, we immediately see that the four derivative terms gives zero, and that the result can be written as:
\begin{equation}
\frac{\overline{\langle\delta\theta\rangle}}{3H}  = - \frac{25}{54}   \frac{1}{a^2H^2}
\overline{\left\langle\varphi_{,i}\varphi^{,i}\right\rangle}  \, .
\label{resultMean}
\end{equation}

Now let us comment on the two derivative term. Symbolically we will call it $k^2\varphi^2$ term.
This contribution was found to be of order $10^{-5}$. We stress now that this is {\it not} a small correction, in the sense that naively one could expect a second order term to be of order $10^{-10}$.
This is the crucial point: the presence of the two powers of $k$ boosts the contribution of five orders of magnitude approximatively at the present epoch. The reason for this is that the scales $k_0$ which have had more time to grow are scales much smaller than the horizon, and so a factor $(k_0/H)^2$ gives a big boost.
We are going to show in detail why this happens, and show that this signals the breakdown of the perturbative expansion.
In fact, 
extending the estimates to analoguous terms that will appear at higher order in perturbation theory, we find that the expansion parameter is of order 1 at late times.

Let us have a look at higher order terms.
Of course, being interested only in the mean, and assuming the variable $\varphi$ gaussianly distributed, only the even orders of perturbation are relevant.
So the next relevant order is the fourth order. For each order the maximum number of spatial derivatives grows by two. As we said the more derivatives we have the bigger is the contribution to the mean.
So at fourth order the highest derivative term is $k^8 \varphi^4$.
As we will show, it is possible to see that such a term again yields a order $1$ possible correction.
At this point one might be tempted to speculate that at fourth order this may not be zero (as in~\cite{Rasanen}). However, the fact that the highest derivative term is a total spatial gradient at second order may not be accidental, and one might expect that at any order in perturbation theory the term with the highest number of derivatives always appear as a spatial gradient. 
So, in order to be conservative, we will not rely on this speculation, and assume that all the terms with the highest number of derivatives are zero at any order.
Also, as we discuss in section~\ref{Newton}, the highest derivative terms are the newtonian terms and it is expected to get zero mean corrections at the newtonian level.

Now, the next term at fourth order is $k^6 \varphi^4$, and by analogy with the second order term $k^2 \varphi^2$, we expect it to be nonzero.
So, generically at order $n$, we assume terms like $k^{2n}\varphi^{n}$ to be zero, while we keep terms like $k^{2n-2}\varphi^{n}$.

For comparison with~\cite{Rasanen}, we stress again that here we are not relying on effects due to UV divergencies (that is to non linear structures), since our considerations involve effects due to scales which are still linear today.  And we do not rely on newtonian terms, but our effect is purely relativistic (see section~\ref{Newton}).

\section{The expansion parameter}

So let us analyze the magnitude of the $k^2\varphi^2$ term.
Assuming the usual flat spectrum for $\varphi$, its value is given by (\cite{KMNR}):
\begin{equation}
\overline{k^2 \varphi^2}=A^2 \frac{a}{a_0} \left( \frac{H \, \Gamma }{h Mpc^{-1}} \right)^2 
\int_0^{\infty} dq \, q \, T^2(q)  \label{k^2}
\end{equation}
where $A$ is the amplitude of the spectrum of $\varphi$ and it is given by
$1.9 \times 10^{-5} /\Omega_M^{(0)}$, where we took into account of the present value of $\Omega_M$ in the normalization~(\cite{Bunn}). Then, $T(q)$ is the Bardeen, Bond,
Kaiser, Szalay (BBKS) transfer
function \cite{BBKS}:  
\begin{eqnarray}
T(q)& =& \frac{\ln\left(1+2.34q\right)}{2.34q}\times \nonumber \\
 && \hspace{-11mm} \times \left[
1+ 3.89q + (16.1q)^2 + (5.84q)^3 +(6.71q)^4\right]^{-\frac{1}{4}} .
\end{eqnarray}
which describes the stagnation effect of the small scale perturbations that entered the horizon during the radiation era.
Of course at small $q$, $T^2(q)\rightarrow 1$, while at large $q$,
$T^2(q)\rightarrow q^{-4}\ln^2\!q$. The integral receives the biggest contribution from the scale at which the integrand has a maximum, which corresponds to ${\cal O}(20-50)$ Mpc.
Here $\Gamma$ is the shape factor, defined for a flat universe in
terms of the baryon fraction $\Omega_B$ and the total value of
$\Omega_M^{(0)}$ as $\Gamma=\Omega_M^{(0)} h
\exp(-\Omega_B-\sqrt{2}h\Omega_B/\Omega_M^{(0)})$. $\Gamma$ is sensitive especially to the value of $\Omega_M^{(0)} h$ and it is decreasing as $\Omega_M^{(0)} h$ increases.
Finally, the integral in eq.(\ref{k^2}) has the numerical value of $2.3\times 10^{-2}$.

For comparison we may estimate a $k^4\varphi^2$ term (although as we said it is not present in the result given by eq.(\ref{resultMean})):
\begin{equation}
\overline{k^4 \varphi^2}=A^2 \left(\frac{a}{a_0}\right)^2 \left( \frac{H \, \Gamma }{h Mpc^{-1}} \right)^4 
\int_0^{\infty} dq \, q^3 \, T^2(q)  
\end{equation}
Such a result would imply a big correction already at second order. In fact the factor $\left( \frac{H \, \Gamma }{h Mpc^{-1}} \right)^4$ is as large as ten or eleven orders of magnitude (depending on the value of cosmological parameters). In addition, the integral is even logaritmically divergent in the ultraviolet, and so gets its main contribution from the small scales. Even putting a cutoff at the nonlinear scales, the integral results in a ${\cal O}(1)$ factor. And finally this correction increases with time twice as fast as the $k^2\varphi^2$ term, becoming big at late times.

Now let us go to estimate the higher orders.
As we said, going to fourth order, we safely assume that the leading term is $k^6\varphi^4$. This is estimated as follows:
$$
\overline{k^6 \varphi^4}=A^4  \left(\frac{a}{a_0}\right)^3 \left( \frac{H \Gamma}{h Mpc^{-1}} \right)^6 
\times Int^3
$$
Here the term $Int^3$ stands for one or more integrals in $dq$. Depending on what is the precise structure of the contraction of indices in the $k^6\varphi^4$ term, it may contain also ultraviolet divergent integrals which would give a larger contribution than the integral that we found at second order, and might be sensitive to the small scales. However, in order to make a conservative estimate we do not rely on such speculations and we just assume the minimal value
$$Int \simeq \int_0^{\infty} dq \, q \, T^2(q) \approx 2.3\times 10^{-2} .$$

\begin{figure}
\includegraphics[width=0.45\textwidth]{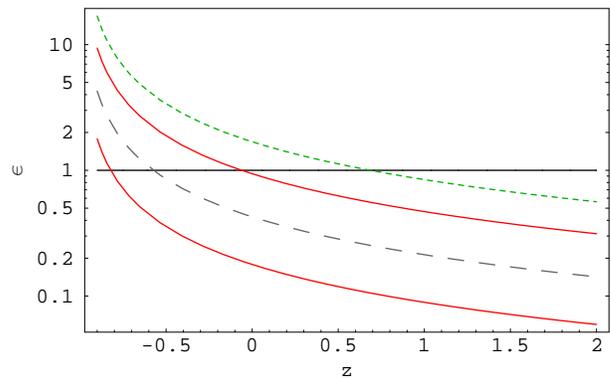}
\caption{\label{figuraCDM}\em 
The expansion parameter $\epsilon$ as a function of the redshift, for different values of the parameters ($h=0.72\pm0.05,\, \Omega_M h^2=0.14\pm0.02,\, \Omega_b h^2=0.024\pm0.001,\, A/\Omega_M=1.9\pm 0.2$, as taken from WMAP only(\cite{WMAP})). The grey dashed line corresponds to the central value, while the red solid lines to $2\sigma$ ranges. Here we used the growth factor as in matter domination. For comparison the green dotted line corresponds to $\Omega_M=1$. The perturbative series breaks down at small $z$ when $\epsilon$ becomes of order 1 (we also extended the plot up to negative $z$ in order to show that at some point the series badly diverges unavoidably.)}
\end{figure}

At the generic order $n$ (where $n$ is even) so we have:
$$
\overline{k^{2n-2} \varphi^{n}} \simeq  A  \epsilon^{n-1}\approx 10^{-5}\epsilon^{n-1}
$$
where $\epsilon$ is the expansion parameter:
$$
\epsilon \equiv \frac{A}{1+z}\left( \frac{H_0 \Gamma}{h Mpc^{-1}} \right)^2 \times Int
$$
The crucial fact is that this expansion parameter becomes of order 1 for small $z$ (the value depends on uncertainties on the cosmological parameters, mainly $\Omega_M$ through $\Gamma$):
\begin{equation}
\epsilon \approx \frac{4 \Gamma^2}{\Omega_M (1+z)} \, ,
\end{equation}
while already for $z\gtrsim 1$, the terms of the series start decreasing, so one may assume that the series converges. The expansion parameter is plotted in fig.\ref{figuraCDM}, for different values of the cosmological parameters.

In a similar way one can also see that the highest $k$ terms (that we set to zero) are of the same order of $k^4\varphi^2$ at small redshifts.

\section{Inhomogeneities and Acceleration}

Now, we may speculate that the fact that the expansion parameter is of order $1$ just around $z\approx 1$ is related to the appearance of acceleration in the observed Universe. If this is the case, this fully solves the coincidence problem. 
Even more: one can see that the time at which $\epsilon$ is of order $1$ depends on the amplitude of the initial fluctuations, so it may appear a coincidence that the initial value of $10^{-5}$ is such that the expansion breaks down right today.
However for any initial amplitude the time at which the series breaks down is always around the time of structure formation, and since we can only live around that time, there is no coincidence at all.

\begin{figure}
\includegraphics[width=0.45\textwidth]{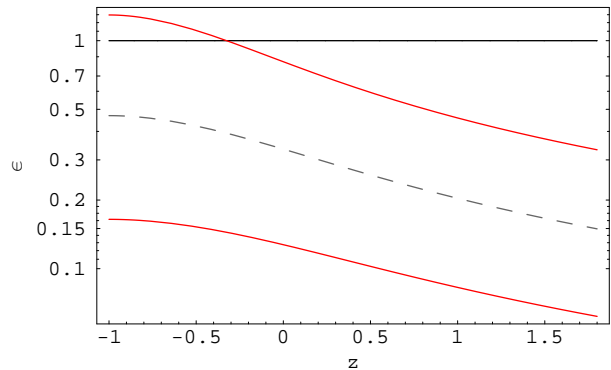}
\caption{\label{figuraLCDM}\em 
The expansion parameter $\epsilon$ as a function of the redshift, for different values of the parameters ($h=0.72\pm0.05,\, \Omega_M h^2=0.14\pm0.02,\, \Omega_b h^2=0.024\pm0.001,\, A/\Omega_M=1.9\pm 0.2$, as taken from WMAP only(\cite{WMAP})). The grey dashed line corresponds to the central value, while the red solid lines to $2\sigma$ ranges. Here we used the growth factor for density perturbations of a $\Lambda CDM$ model~{\cite{Carroll}}, in order to show that $\epsilon$ is prevented to grow and it stabilizes at late times, if the Universe starts accelerating.}
\end{figure}

At this point, so, let us assume that the Universe starts accelerating as a result of the divergence of the series. This, in turn, stops the growth of the density perturbations, and so the expansion parameter itself is prevented to grow and it stabilizes around the value that it takes when the effect starts to be important (as one can see in fig.~(\ref{figuraLCDM})).

It has been pointed out in~\cite{Seljak} and~\cite{Flanagan} that a deceleration parameter defined in terms of a local expansion rate can never become negative in absence of vorticity and in absence of negative pressure fluids. However this result does not apply here, since the expansion we are considering is not $\theta$, but is $\langle \theta \rangle$. When trying to derive the same result to this quantity, one can see that new negative terms appear due to the time derivative of the $\sqrt{\gamma}$, which is present in the definition of the spatial average.

In order to show this explicitly we need to introduce the expansion tensor, which reads
\begin{equation}
\theta_{\phantom{i} j}^{i}=\frac{1}{2 a^2} \gamma^{ik} \frac{d}{dt}(a^2 \gamma_{kj})  \label{thetaij} \, .
\end{equation}
This can be also decomposed into its trace (which is $\theta\equiv \theta_{\phantom{i} i}^{i}$) and a traceless tensor (the shear tensor):
\begin{equation}
\theta_{\phantom{i} j}^{i}\equiv \frac{1}{3}\theta \delta_{\phantom{i} j}^{i}+\sigma_{\phantom{i} j}^{i} \label{sigma}  \, .
\end{equation}

Once we have defined the expansion rate as $\langle \theta \rangle/3$, its deceleration parameter is given by:
\begin{equation}
q\equiv -1-3\frac{\frac{d}{dt}\langle \theta \rangle}{\langle \theta \rangle^2}
\end{equation}

The 
Raychaudhuri equation gives: 
\begin{equation}
\frac{d \theta} {d t}=-\frac{\theta^2}{3}-\sigma^2
-4\pi G (\rho+3p)
\end{equation}
where $\sigma^2 \equiv\sigma_{\phantom{i} j}^{i} \sigma_{\phantom{i} i}^{j}$. And where we assumed zero pressure and zero vorticity.
Another useful relation is obtained taking the determinant of eqs.~(\ref{sigma}) and~(\ref{thetaij}), which gives $\frac{1}{2 \gamma} \frac{d\gamma}{dt}=\frac{\theta}{3}+\sigma$.

Using these relations it is straightforward to obtain the following:
\begin{eqnarray}
q &=& 3
\frac{ \langle \sigma^2 \rangle }{ \langle \theta \rangle^2}  + 
12 \pi G \frac{\langle\rho\rangle}{\langle \theta \rangle^2} + \nonumber  \\
&-& \frac{1}{2}\left( \frac{\langle \theta^2 \rangle-\langle \theta \rangle^2}{\langle \theta \rangle^2} \right)-
\frac{3}{2}\left( \frac{\langle \theta \sigma \rangle-\langle \theta \rangle\langle \sigma \rangle }{\langle \theta \rangle^2} \right) ,
\end{eqnarray}
which is an exact equation, valid in general.
The first line corresponds to what was derived in \cite{Seljak}, while the second line arises due to the presence of $\sqrt{\gamma}$ in the averages.

This clearly shows that there are negative contributions and so
 when the corrections overcome the background value, the deceleration parameter can change sign.

%%%%%%%%%%%%%%%%%%%%%%%%%%%%%%%%%%%%%%%%%%%%%%%%%%%%%%%%%%%%%%%%%%
%%%%%%%%%%%%%%%%%%%%%%%%%%%%%%%%%%%%%%%%%%%%%%%%%%%%%%%%%%%%%%%%%%

%%%%%%%%%%%%%%%%%%%%%%%%%%%%%%%%%%%%%%%%%%%%%%%%%%%%%%%%%%%%%%%%%%
%%%%%%%%%%%%%%%%%%%%%%%%%%%%%%%%%%%%%%%%%%%%%%%%%%%%%%%%%%%%%%%%%%
%%%%%%%%%%%%%%%%%%%%%%%%%%%%%%%%%%%%%%%%%%%%%%%%%%%%%%%%%%%%%%%%%%
%%%%%%%%%%%%%%%%%%%%%%%%%%%%%%%%%%%%%%%%%%%%%%%%%%%%%%%%%%%%%%%%%%

%%%%%%%%%%%%%%%%%%%%%%%%%%%%%%%%%%%%%%%%%%%%%%%%%%%%%%%%%%%%%%%%%%
%%%%%%%%%%%%%%%%%%%%%%%%%%%%%%%%%%%%%%%%%%%%%%%%%%%%%%%%%%%%%%%%%%
%%%%%%%%%%%%%%%%%%%%%%%%%%%%%%%%%%%%%%%%%%%%%%%%%%%%%%%%%%%%%%%%%%

\section{Newtonian and Post-Newtonian terms} \label{Newton}
 
It is very interesting to note that (see~\cite{MatTerr}) the highest $k$ terms (that we neglected) are also the terms with the highest power of $a\propto \tau^2$ (we are using here the simple growth factor and background evolution of a pure matter dominated Universe). Now, in an expansion in powers of the speed of light $c$, each power of $\tau$ is accompanied by a power of $c$. So the highest $k$ terms are the  newtonian terms (\cite{MatTerr}).

We stress instead that the terms that we considered here, instead, are the post-newtonian ones. 

The following example is quite illuminating in order to understand why the highest $k$ terms should give zero effect. Consider a homogeneous universe filled with many spherical inhomogeneous regions: outside these regions matter is uniformly distributed with energy density $\rho_0$, and inside these regions matter density is inhomogeneously distributed (but still maintaining the spherical symmetry and in such a way that the total amount of matter inside a sphere is equal to the amount of a same-sized sphere with homogeneous matter density $\rho_0$)\footnote{We thank N.Afshordi for pointing out this example.}.

Now, in newtonian gravity, we can use the superposition principle and consider separately the effect of a single sphere. By Gauss' theorem the only thing that matters is the total amount of matter inside the sphere, so the average expansion rate over a region that contains spheres is the same one of a perfectly homogenous universe with matter density $\rho_0$.

The only way in order to guarantee that the effect is zero in this example is that the Newtonian terms be total spatial gradients. In fact in this case one can rewrite the spatial average as a flux of a vector field over a surface. Taking the surface in such a way that it does not intersect any sphere, this gives zero because the vector field in the homogeneous region is zero.

This shows that Newtonian terms have to be surface terms, and in order to have an effect one has to consider the purely relativistic terms (and the post-newtonian approximation should be accurate enough).

So, what we are claiming is that the series (in an expansion in the smallness of the metric elements) of all the post-newtonian terms is diverging, and we may speculate that the resummation of all these terms may give rise to an accelerating Universe. Or in other words it may give rise to an effective theory in which the Friedmann equations are still valid, but with an effective $\Lambda$ term. In order to prove this kind of ``phase transition'' at redshift of order 1, so, one needs a treatment of the postnewtonian terms, which does not assume that the metric elements are small.

\section{Conclusions}

We have shown that, even being very conservative, the contribution of the inhomogeneities to the average expansion rate becomes perturbatively out of control at low redshifts. This leads to the conclusion that the usual Friedmann equation cannot be trusted at late time.

So, instead of using the naive homogeneous model to fit the data, one should to take a nonperturbative approach to Einstein equations in a inhomogeneous Universe, before talking of Dark Energy.

We suggest that the failure of this equation might be the reason why we need to introduce mysterious components in order to fit the data. We speculate that the resummation of the terms of the diverging series (which may act as a negative pressure component) might lead to an effective Friedmann equation in which the series is resummed, giving an effective ``Dark Energy''.
Once this effect is taken into account, the perturbations themselves stop growing (as happens in a $\Lambda CDM$ model), and we speculate that the effect itself may be stabilized, reaching a fixed point, in which it might even be possible to predict the effective $\Omega_M\approx0.3$.

\medskip

%%%%%%%%%%%%%%%%%%%%%%%%%%%%%%%%%%%%%%%%%%%%%%%%%%%%%%%%%%%%%%%%%%%%%%%%%
%%%%%%%%%%%%%%%%%%%%%%%%%%%%%%%%%%%%%%%%%%%%%%%%%%%%%%%%%%%%%%%%%%%%%%%%%
\section*{Acknowledgments}
%%%%%%%%%%%%%%%%%%%%%%%%%%%%%%%%%%%%%%%%%%%%%%%%%%%%%%%%%%%%%%%%%%%%%%%%%
%%%%%%%%%%%%%%%%%%%%%%%%%%%%%%%%%%%%%%%%%%%%%%%%%%%%%%%%%%%%%%%%%%%%%%%%%
We would like to thank Rocky Kolb, Sabino Matarrese and Toni Riotto for previous work and discussions on the subject. It is a pleasure to thank also  Niayesh Afshordi, Thorsten Battefeld, Robert Brandenberger, Paul Chouha, Fabrizio Di Marco, Balaji Katlai, Natalia Shuhmaher and Alberto Vallinotto for useful comments and discussions.

%%%%%%%%%%%%%%%%%%%%%%%%%%%%%%%%%%%%%%%%
%%%%%%%%%%%%%%%%%%%%%%%%%%%%%%%%%%%%%%%%

%%%%%%%%%%%%%%%%%%%%%%%%%%%%%%%%%%%%%%%%
%%%%%%%%%%%%%%%%%%%%%%%%%%%%%%%%%%%%%%%%

%%%%%%%%%%%%%%%%%%%%%%%%%%%%%%%%%%%%%%%%
%%%%%%%%%%%%%%%%%%%%%%%%%%%%%%%%%%%%%%%%
\end{document}